\documentclass[aps, prl, superscriptaddress, twocolumn, floatfix]{revtex4-2}
\usepackage{graphicx, epsfig, color}
\usepackage{amssymb}
\usepackage{amsmath}
\usepackage{hyperref}
\usepackage{bm}
\begin{document}
\title{Adiabatic State Preparation in a Quantum Ising Spin Chain}
\author{Sooshin Kim}\email{sooshinkim@postech.ac.kr}\altaffiliation[current address: ]{Basic Science Research Institute, Pohang University of Science and Technology (POSTECH), Pohang 37673, Korea}
\author{Alexander Lukin}
\author{Matthew Rispoli}
\author{M. Eric Tai}
\affiliation{Department of Physics, Harvard University, Cambridge, Massachusetts 02138, USA}
\author{Adam M. Kaufman}
\affiliation{JILA, University of Colorado and National Institute of Standards and Technology, and Department of Physics, University of Colorado, Boulder, Colorado 80309, USA}
\author{Perrin Segura}
\author{Yanfei Li}
\author{Joyce Kwan}
\affiliation{Department of Physics, Harvard University, Cambridge, Massachusetts 02138, USA}
\author{Julian L\'eonard}
\affiliation{Vienna Center for Quantum Science and Technology (VCQ), Atominstitut, TU Wien, 1020 Vienna, Austria}
\author{Brice Bakkali-Hassani}
\author{Markus Greiner}
\affiliation{Department of Physics, Harvard University, Cambridge, Massachusetts 02138, USA}

\begin{abstract}
We report on adiabatic state preparation in the one-dimensional quantum Ising model using ultracold bosons in a tilted optical lattice. We prepare many-body ground states of controllable system sizes and observe enhanced fluctuations around the transition between paramagnetic and antiferromagnetic states, marking the precursor of quantum critical behavior. Furthermore, we find evidence for superpositions of domain walls and study their effect on the many-body ground state by measuring the populations of each spin configuration across the transition. These results shed new light on the effect of boundary conditions in finite-size quantum systems.
\end{abstract}

\maketitle

Quantum Ising spin models are textbook examples for the study of many-body phenomena and quantum phase transitions \cite{2008Sachdev, 1994Auerbach_Book}. In recent years, various platforms have performed quantum simulation of these models, including trapped ions \cite{2021Monroe} and Rydberg atom arrays \cite{2018Schauss, 2021Ebadi, 2021Scholl}. It was originally suggested in Refs.\,\cite{2002Sachdev_n_Girvin, 2003Krivnov} that ultracold bosons in a tilted optical lattice could realize an Ising model, which was later demonstrated experimentally \cite{2011Simon_n_Greiner,  2013Nagerl}. Access to the ground state for large systems described by these models can be difficult, as the many-body gap is  generically small and closes when a phase transition is crossed. On the other hand, recent experiments have motivated the use of small systems with well-controlled size as a bottom-up approach to quantum simulation \cite{2022Sompet, 2023Leonard}, for which energy gaps remain significant and therefore allow for adiabatic quantum state engineering.

In this Letter, we perform adiabatic ramps in a one-dimensional (1D) tilted optical lattice and probe ground states of a quantum Ising spin model with both transverse and longitudinal fields \cite{2002Sachdev_n_Girvin, 2011Simon_n_Greiner}. We benefit from a dynamical scale on the order of the tunneling amplitude $t$, which is larger than the usual super-exchange interaction $t^2 / U$, where $U$ is the on-site interaction energy. Our work distinguishes itself through the use of full-counting statistics to map out the paramagnetic-to-antiferromagnetic (PM-to-AFM) quantum phase transition and confirm the fidelity of our preparation scheme. Furthermore, by controlling the spin chain size, we shed light on the role of boundary conditions in our finite-size system through the presence of domain walls and the build-up of entanglement across the chain \cite{2015Owerre, 2020Maric, 2021Maric}. 

\begin{figure}
	\centering
	\includegraphics{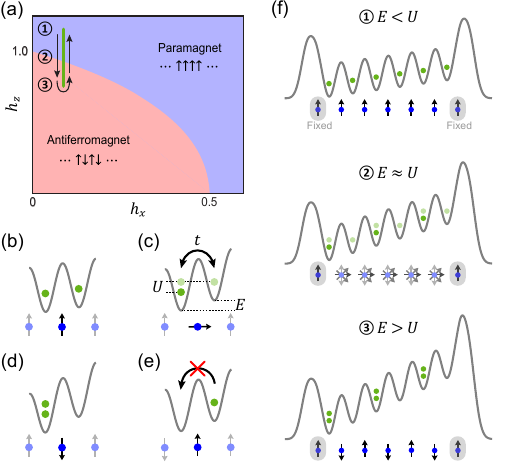}
	\caption{Experimental protocol. (a) Ground-state phase diagram of a 1D Ising chain with antiferromagnetic interactions and both longitudinal ($h_z$) and transverse ($h_x$) fields~\cite{2003Krivnov}. In our experiment, $h_z$ is swept along the green line, while the transverse magnetic field $h_x = 0.086(2)$ is kept constant. (b) In a weakly-tilted lattice, an atom remaining on its original site is mapped onto state $|{\uparrow}\rangle$. (c) When the tilt per lattice site $E$ approaches the on-site interaction energy $U$, the atom delocalizes over the two neighboring lattice sites. (d) When the lattice is more strongly tilted, an atom that has hopped to the next site is represented by state $|{\downarrow}\rangle$. (e) In the $h_z$ range considered here, an atom cannot hop to a vacant neighboring site. (f) After initializing the system in the PM state, we sweep $E$ across the $U$ resonance and then rewind the ramp over the same duration.}
	\label{figure1}
\end{figure}

Our system is described by the following Hamiltonian \cite{2002Sachdev_n_Girvin, 2011Simon_n_Greiner}
\begin{equation}
\label{eq:Ising}
	H = J \sum_j \left( S_j^{z} S_{j + 1}^{z} -  h_z S_j^{z} - h_x S_j^{x} \right),
\end{equation}
where $S_j^{z, x}$ are the components of a spin-$1/2$ located at site $j$, $J > 0$ is the nearest-neighbor antiferromagnetic coupling strength, and $h_z$ (resp. $h_x$) is the longitudinal (resp. transverse) component of the external magnetic field. Fig.\,\ref{figure1}(a) summarizes the phase diagram of this Ising Hamiltonian: strong magnetic fields tend to align the spins and tip the system into the PM phase, while in the opposite case, the spins anti-align and the system chooses the AFM phase \cite{2003Krivnov}. A second-order phase transition separates the two phases, except at $(h_x, h_z) = (0, 1)$ for which the transition is first order.

Our experiments start with a two-dimensional Mott insulator of $^{87}$Rb atoms at unity filling, from which we isolate a single one-dimensional chain of $L$ sites using our high-resolution imaging system and site-resolved optical potentials \cite{IsingSM}. In the subsequent steps, we apply a linear potential along the chain such that our Bose-Hubbard system can be mapped onto the spin model \eqref{eq:Ising}, as demonstrated experimentally in Ref.\,\cite{2011Simon_n_Greiner}. After imposing an initial energy offset $E = 0.705(9) \times U$ between adjacent lattice sites, while the on-site interaction energy is set to $U = h \times 404(4)$~Hz, we reduce the lattice depth along the chain to restore tunneling with an amplitude $t = 0.030(1) \times U$ \cite{IsingSM}. Here, $h$ is the Planck constant.

Deep in the Mott-insulating regime $t \ll U$, and when $E \ll U$, an atom located at site $j$ cannot hop to a neighboring site. We map this configuration onto the eigenstate $|{\uparrow}\rangle$ of an operator $S_j^z$ for a pseudo-spin living on the bond of the lattice, see Fig.\,\ref{figure1}(b). By adiabatically increasing $E$ towards $U$, the two configurations shown in Fig.\,\ref{figure1}(c) become degenerate and the atom delocalizes over the two neighboring sites. As the tilt $E$ is further increased, the initial configuration of Fig.\,\ref{figure1}(b) is energetically suppressed and the atom now occupies the lower site, represented by the pseudo-spin state $|{\downarrow}\rangle$, as shown in Fig.\,\ref{figure1}(d). The intermediate configuration in Fig.\,\ref{figure1}(c) thus corresponds to an equal superposition of spin states $|{\uparrow}\rangle$ and $|{\downarrow}\rangle$, i.e.~to the eigenstate of a transverse spin component $S_j^x$. Fig.\,\ref{figure1}(e) goes beyond the previous simplification based on a double-well picture and shows a situation in which the neighboring site is vacant because its original atom has already hopped to the next site on the left. Then, even when $E = U$, an atom located on the right site is not energetically allowed to hop to the empty site, leading the system to arrange itself as a charge density wave consisting of alternating empty and doubly-occupied sites, or as an antiferromagnet in the spin language. Overall, the mapping between the Bose-Hubbard model and the spin model \eqref{eq:Ising} is realized with $J \simeq U$ for the antiferromagnetic coupling strength, and $h_z = 1 - (E - U)/J$ and $h_x = 2^{3/2} \, t / J$ for the longitudinal and transverse magnetic field components, respectively. In the regime $0 < h_x \ll 1$ relevant for our experiments, the slowest dynamical scale is therefore given by the tunneling amplitude $t$, which is much faster than the typical super-exchange interaction $t^2 / U$ \cite{2015Hart, 2017Mazurenko, 2021Pan, 2021Chung}. In addition, we project sharp potential walls at both ends of the chain to confine the dynamics to a finite region. The presence of hard walls imposes fixed boundary conditions on the spins: since the leftmost atom cannot hop further to the left and no atom can hop from the right into the rightmost site, this situation is equivalent to having $|{\uparrow}\rangle$--spins coupled at both ends of the chain, see Fig.\,\ref{figure1}(f). Consequently, our system is made of $N_{\mathrm{s} } = L + 1$ spins in total.

After initializing the system in the paramagnetic regime with $h_z = 1.30(1)$, we linearly tune the longitudinal field $h_z$ to $0.73(1)$ within $250$~ms, such that the atoms adiabatically follow the ground state of $H$ as the system crosses the region around the quantum critical point (QCP) whose numerically predicted location in the thermodynamic limit \cite{2003Krivnov} is $h_z^{\mathrm{c} } = 0.943(1)$ for $h_x = 0.086(2)$ in our experiment. At variable points of the ramp, we rapidly increase the lattice depth to freeze the dynamics and expand the atoms along tubes perpendicular to the chain~\cite{2016Kaufman_n_Greiner}. This allows us to count the number of atoms in each lattice site through fluorescence imaging, thereby circumventing the limitations set by parity-projection in multiply-occupied sites. Alternatively, in order to check the adiabaticity of the entire process, we rewind the ramp back to its initial value before imaging the atoms. We postselect realizations whose number configuration can be mapped onto the pseudo-spin Hilbert space \cite{IsingSM}, and compare our experimental results with the properties of the instantaneous ground states of $H(t)$.

\begin{figure}
	\centering
	\includegraphics{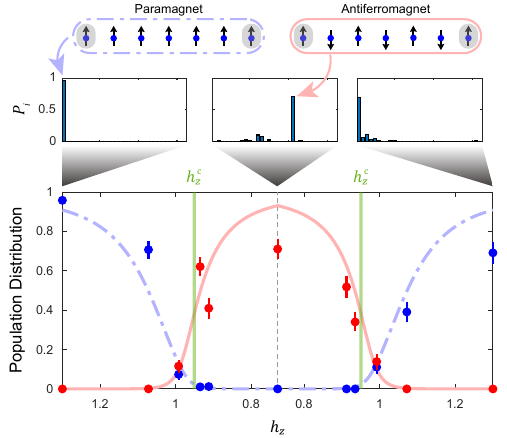}
	\caption{Observing the Ising quantum phase transition between paramagnetic and antiferromagnetic states: Probability of observing different spin configurations in a chain with length $N_{\mathrm{s} } = 7$. At $h_z > h_z^{\mathrm{c} } = 0.943(1)$, the paramagnetic (PM) configuration is the most represented state, while the antiferromagnetic (AFM) configuration becomes dominant when $h_z < h_z^{\mathrm{c} }$. The histograms show the populations of all $2^{N_\mathrm{s} - 2} = 32$ spin configurations at three points of the ramp. The position of the QCP expected in the thermodynamic limit is shown as a green vertical line. The blue and red lines show the prediction of exact diagonalization calculations without any free parameters. Error bars denote the 1$\sigma$ statistical errors.}
	\label{figure2}
\end{figure}

We now turn to the observation of the PM--AFM quantum phase transition in our system. In Fig.\,\ref{figure2}, we present the evolution of the population distribution for a system of size $N_{\mathrm{s} } = 7$ during our experimental ramp, in particular the spin configuration histograms at three different times during the ramp. As long as the longitudinal field $h_z$ is far from the QCP, there is no significant change in the population distribution. When we increase the tilt through the QCP, the PM state population is transferred to the AFM state. Conversely, the opposite phenomenon occurs when the ramp is rewound over time, confirming the adibaticity of the process. The transition itself is smoothed and slightly offset from the thermodynamic prediction $h_z =  0.943(1)$ due to the finite size of our system.

\begin{figure}
	\centering
	\includegraphics{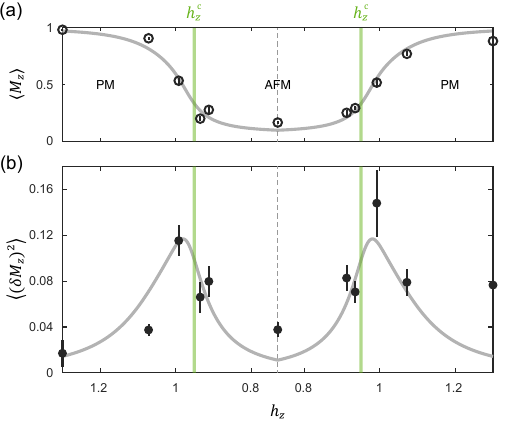}
	\caption{Enhanced fluctuations across the quantum phase transition. (a) Evolution of the mean magnetization and (b) its fluctuation, illustrating the precursor of quantum critical behavior when the phase transition is crossed. The position of the QCP expected in the thermodynamic limit is shown as a green vertical line. The solid lines show the prediction of exact diagonalization calculations without any free parameters. Error bars denote the 1$\sigma$ statistical errors, obtained using bootstrap in (b).}
	\label{figure3}
\end{figure}

Generically, phase transitions are driven by collective fluctuations in the components of the system at the critical point~\cite{2017Tauber, 2018Si}. We investigate the precursor of this critical behavior around the QCP by extracting the mean magnetization $\langle M_z \rangle$ and its fluctuations \cite{IsingSM}. As shown in Fig.\,\ref{figure3}(a) for the case $N_{\mathrm{s} } = 7$, the mean magnetization $\langle M_z \rangle$ remains close to unity in the PM regime and falls to almost zero in the AFM regime, being consistent with numerical expectation. In parallel, the magnetization shows enhanced fluctuations at the QCP, both in the forward and reverse parts of the ramp [Fig.\,\ref{figure3}(b)]. This observation shows that enhanced quantum fluctuations of the spin chain occur at the QCP, a hallmark of quantum criticality.

\begin{figure}
	\centering
	\includegraphics{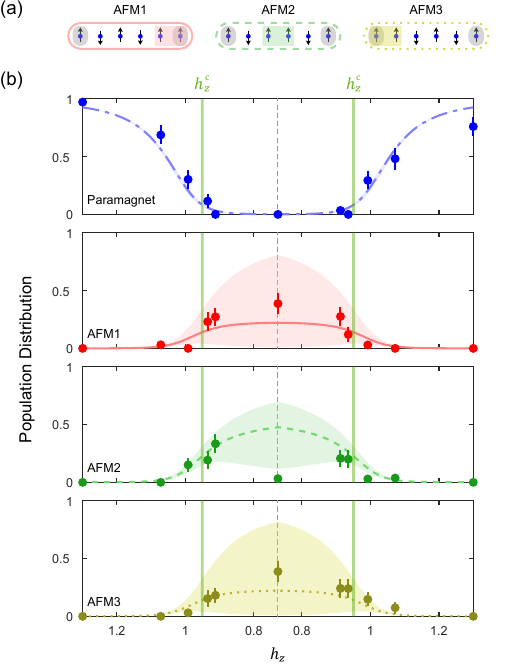}
	\caption{Probing an antiferromagnetic chain with frustrated boundary conditions. (a) When $h_x$ = 0, the ground state of a finite chain with even $N_{\mathrm{s} }$ is $N_{\mathrm{s} } / 2$ times degenerate in the AFM regime. Each classical configuration contains a domain wall made of two adjacent $|{\uparrow}\rangle$--spins. (b) Population distribution for a system size $N_{\mathrm{s} } = 6$ as we sweep $h_z$. Multiple AFM configurations appear at the QCP and coexist in the AFM regime. The position of the QCP expected in the thermodynamic limit is shown as a green vertical line. The other colored lines show the prediction of exact diagonalization calculations without any free parameters, while the shaded areas show the uncertainty for the expected potential disorder in the experiment \cite{IsingSM}.}
	\label{figure4}
\end{figure}

We now consider a chain made of an even number of spins $N_s$. Due to the fixed boundary conditions imposed on the spin chain, the spectrum of the system dramatically depends on the parity of $N_{\mathrm{s} }$. Let us first consider the case $h_x = 0$ for simplicity: While there is a single AFM ground state when $N_{\mathrm{s} }$ is odd, the ground state becomes degenerate in the AFM regime when $N_{\mathrm{s} }$ is even. More precisely, each state of the $N_{\mathrm{s} } / 2$--degenerate manifold contains exactly one domain wall made of adjacent $|{\uparrow}\rangle$--spins, see Fig.\,\ref{figure4}(a). Such frustrated boundary conditions, where the local antiferromagnetic order cannot propagate from one end of the chain to the other, have been studied in Refs.\,\cite{2015Owerre, 2020Maric, 2021Maric}. Intuitively, this domain-wall excitation behaves as a single particle without any kinetic energy, i.e. with an infinite mass. A finite but small $h_x$ perturbatively lifts this degeneracy by conferring a finite mass to these excitations \cite{IsingSM}. It also allows us to perform adiabatic ramps through a non-degenerate ground state. In the minimal settings of two pseudo-spins with two boundary spins ($N_{\mathrm{s} } = 4$ in total), the state realized in the AFM phase and small $h_x$ is a maximally-entangled Bell pair. Consequently, a chain with an even number of spins gives access to non-classical superpositions of spin configurations.

Fig.\,\ref{figure4}(b) shows the transfer of the PM population in three competing AFM configurations for a chain made of $N_{\mathrm{s} } = 6$ spins. Different AFM populations appear around the QCP and coexist in the AFM regime. When we rewind the ramp, these AFM populations collapse and the PM population is restored. This reversible behavior indicates that during the ramp, a superposition of competing AFM configurations is realized while maintaining adiabaticity. Such a superposition implies the presence of entanglement between distant spins, which is also confirmed in our experiment by extracting the single-site entropy \cite{IsingSM}. In Fig.\,\ref{figure4}(b), we attribute the discrepancy between theory and experiments to residual on-site potential disorder which can significantly affect the population distributions among the quasi-degenerate states in the AFM regime, as indicated by the shaded areas which were obtained by sampling random disorder distributions.

As a conclusion, we experimentally realized a continuous paramagnetic-to-antiferromagnetic quantum phase transition using a 1D tilted Bose-Hubbard system. We showed the reversible transfer of population between the two phases and investigated the critical behavior of the system indicated by increased magnetization fluctuations. Most remarkably, we created a domain wall superposition by making use of frustrated boundary conditions. Important technical improvements with respect to Ref.~\cite{2011Simon_n_Greiner} include the control of the system size, the implementation of full-counting statistics, and the use of post-selection. The main challenges of our approach are non-adiabatic processes and potential disorder, especially harmful in the case of larger system sizes and in the presence of frustrated boundary conditions. These limitations could be addressed in the near future respectively by characterizing and correcting optical potential disorder and via counter-diabatic drives for faster ramps \cite{2017Sels}, yielding enhanced fidelities for the ground-state preparation and providing a route to larger system sizes. 

It has been suggested recently that an antiferromagnetic spin chain with periodic boundary conditions can show a parity-dependent quantum phase transition, suggesting that boundary conditions can dramatically affect the behavior of a macroscopic system \cite{2022Shaikh}. Alternatively, our approach that uses ultracold atoms in tilted optical lattices can be generalized to higher dimension to yield longitudinal density-wave ordering with transverse superfluidity, and to other lattice geometries to generate non-trivial entanglement of pseudo-spin degrees of freedom \cite{2011Pielawa}. Finally, strongly tilted optical lattices can give access to non-ergodic behaviors due to Hilbert space fragmentation \cite{2023Kohlert} and to new phases of matter emerging from kinetic constraints \cite{2023Zechmann, 2022Lake}.

\begin{acknowledgments}
We acknowledge helpful discussions with S.~Choi, J.~H.~Han, P.~Hauke, B.~Kang, S.~Kang, R.~Vatr\'e, and K.~Wang. This work was supported by grants from the National Science Foundation, the Gordon and Betty Moore Foundations EPiQS Initiative, an Air Force Office of Scientific Research MURI program, and an Army Research Office MURI program. J.~L. acknowledges support by the FWF grant no. Y-1436 and the FFG grant no. F0999896041.
\end{acknowledgments}

M.~G. is a cofounder and shareholder of QuEra Computing. All other authors declare no competing interests.

\bibliography{biblio}

\appendix

\clearpage

\section{SUPPLEMENTAL MATERIAL}

\subsection{Calibration of Bose-Hubbard parameters}

The tunneling amplitude $t = h \times 12.3(3)$~Hz is extracted using one-dimensional quantum walks in a flat lattice, while the on-site interaction energy $U = h \times 404(4)$~Hz and the variable gradient energy $E$ are calibrated using photon-assisted tunneling in a tilted lattice. These calibration measurements have been described in detail in Ref.\,\cite{2019Lukin_n_Greiner}.

\subsection{Experimental sequence}

\begin{figure}
	\centering
	\includegraphics{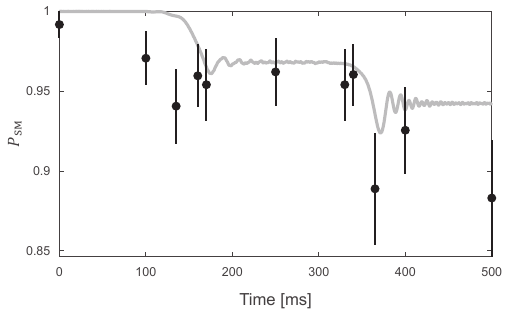}
	\caption{The postselection rate $P_\mathrm{SM}$ (black dot) is obtained by filtering in atom number configurations that can be mapped onto spin states among the snapshots with the correct total atom numbers. The grey solid line shows the weight of the many-body wave function that can be projected onto the spin Hilbert space, after numerically integrating the Schr\"odinger equation for the tilted Bose-Hubbard Hamiltonian $H_\mathrm{B}$. Error bars denote the 1$\sigma$ statistical errors.}
	\label{FigureS:Postselection_Rate}
\end{figure}

Our experiments start with a two-dimensional Mott insulator at unity filling. We prepare $^{87}$Rb atoms in the Zeeman state $|F = 1, m_F = -1 \rangle$ and load them in a single layer of a blue-detuned square optical lattice with lattice constant $a = 680$~nm and $43\,E_\mathrm{r}$ lattice depth, where $E_\mathrm{r} = h \times 1.24$~kHz is the corresponding recoil energy, and $h$ is the Planck constant. Through our microscope setup with sub-lattice spacing resolution, we project a tailoring potential using a digital micromirror device (DMD) to isolate an $L$-site chain of atoms with high fidelity \cite{2009Bakr_n_Greiner, 2016Zupancic_n_Greiner}. The DMD also allows us to generate hard walls at both ends of the chain, which provide a box-like confinement for the 1D chain in the following steps. Next, we apply a linear magnetic field gradient to impose a tilted energy offset $E$ per lattice site. Initially, we set this gradient to $E = h \times 285(1)~\textrm{Hz}~ > U/2$, which limits the contribution of second-order hopping processes that incorporate states outside of the prescribed basis for a faithful mapping to the spin model. Subsequently, the lattice depth is reduced to 16~$E_\mathrm{r}$ within 5~ms, which restores tunneling with amplitude $t = h \times 12.3(3)$~Hz and yields an on-site interaction energy $U = h \times 404(4)$~Hz. Then, we linearly increase $E$ from $h \times 285(1)$~Hz to $h \times 515(2)$~Hz within 250~ms and afterward lower it down to the initial value $h \times 285(1)$~Hz for another 250~ms. At a variable point of the ramp, we freeze the dynamics by increasing the lattice depth to $43\,E_\mathrm{r}$ within 2~ms. The atoms are expanded along the orthogonal direction of the chain to perform a site-resolved atom number measurement through fluorescence imaging \cite{2016Kaufman_n_Greiner}. We repeat the experiment 180 times for each ramp time shown in the main text. We subsequently post-select realizations that contain the correct total number of atoms, yielding $66(4)\%$ post-selection rates for short ramp times and down to $42(4)\%$ for the longest ramp times, for a chain with $L = 6$ sites. We also filter in the snapshots whose number configurations can be mapped to the spin Hilbert space, as shown in Fig.\,\ref{FigureS:Postselection_Rate}.

\subsection{Mapping onto the spin model $H$}

The Bose-Hubbard Hamiltonian with a linear potential can be written as
\begin{eqnarray}
	H_{\rm B} &=& - t \sum_j (a_{j}^{\dag} a_{j + 1} + \mathrm{h.c.} ) + \frac{U}{2}\sum_i n_{j}(n_{j} - 1) \nonumber\\
	&+& E \sum_j j \, n_j,
	\label{eq:H}
\end{eqnarray}
where $a_j^{\dag}$ ($a_j)$ is the bosonic creation (annihilation) operator at site $j$, $n_j = a_j^{\dag} a_j$ is the number operator, $t$ is the nearest-neighbor tunneling amplitude, $U$ is the on-site interaction energy, and $E$ is the potential offset between adjacent lattice sites, analogous to a homogeneous electric field for charged particles. As shown in Refs.\,\cite{2002Sachdev_n_Girvin, 2011Simon_n_Greiner}, the tilted Bose-Hubbard model~\eqref{eq:H} can be mapped onto a one-dimensional chain of pseudo-spins~\eqref{eq:Ising}, with $J \simeq U$ for the antiferromagnetic coupling strength, and longitudinal and transverse magnetic field components $h_z = 1 - (E - U)/J$ and $h_x = 2^{3/2} t / J$, respectively. In the regime $0 < h_x \ll 1$ relevant for our experiments, the quantum critical point is approximately located at $h_z = 1 - 0.66\,h_x$. 

For a finite chain of $L$ sites at unity filling, the dimension of the Hilbert space for the Bose-Hubbard Hamiltonian is given by $D_\mathrm{B} = (2L - 1)! / [ L! \, (L - 1)!] $, such that $D_\mathrm{B} = 126$ for $L = 5$, and $D_\mathrm{B} = 462$ for $L = 6$. The dimension of the Hilbert space for the corresponding spin chain made of $N_{\mathrm s} = L + 1$ spins is $D_\mathrm{S} = 2^{N_\mathrm{s} - 2} = 16$ for $L = 5$, and $D_\mathrm{S} = 32$ for $L = 6$. Every spin state configuration can be mapped onto the Bose-Hubbard Hilbert space, but some configurations lie outside of the energy interval which is well described by our mapping. In particular, this is the case of all states that contain adjacent $|{\downarrow}\rangle$--spins. Conversely, some Fock space configurations cannot be mapped onto the spin basis and have been filtered out for producing the data presented in the main text. As shown in Fig.\,\ref{FigureS:Postselection_Rate}, these states correspond experimentally to a small fraction $< 15\%$ of the realizations that contain the correct total atom number.

\subsection{Numerics}

\begin{figure}
	\centering
	\includegraphics{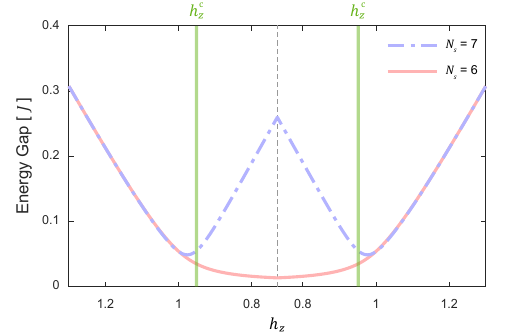}
	\caption{Energy gap between the ground state and the first excited state of $H$ as a function of $h_z$. The transverse field $h_x = 0.086$ is chosen to match our experimental parameters. The position of the QCP expected in the thermodynamic limit is shown as a green vertical line. There is a local minimum at the QCP when $N_{\mathrm s}$ is odd, while for an even $N_{\mathrm s}$, the gap monotonically decreases as $h_z$ gets smaller.}
	\label{FigureS:Gap}
\end{figure}

\begin{figure}
	\centering
	\includegraphics{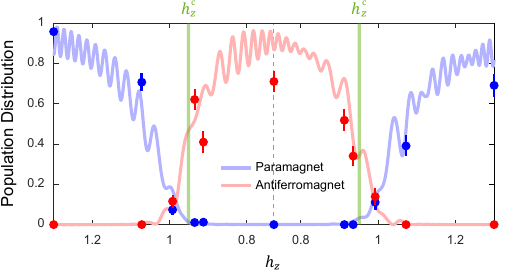}
	\caption{Evolution of the population distributions for a tilted Bose-Hubbard system with length $L = 6$ (which corresponds to the system size of $N_\mathrm{s} = 7$ spins). The position of the QCP expected in the thermodynamic limit is shown as a green vertical line. Solid lines are computed by integrating the time-dependent Hamiltonian $H_{\rm B}(t)$. Experimental data points are the same as shown in Fig.\,\ref{figure2}.}
	\label{figure6}
\end{figure}

We perform exact-diagonalization (ED) calculations to determine the ground state of the spin Hamiltonian $H$ defined in Eq.~\ref{eq:Ising} and to extract expectation values for the different observables presented in the main text. In addition, we show in Fig.~\ref{FigureS:Gap} the energy gap between the ground state and the first excited state of $H$ as a function of $h_z$, which dictates the timescale requirements for our adiabatic preparation scheme. We also simulate numerically the experiment by integrating the Schr\"odinger equation with a time-dependent Bose-Hubbard Hamiltonian $H_{\rm B}(t)$. We trotterize the time evolution and use the Krylov-subspace method to efficiently compute the action of the evolution operation at each time step \cite{1998Sidje}. Finally, we project out the Fock states that cannot be mapped onto the spin basis before extracting any observable. In Figure~\ref{figure6}, we show the experimental data of Fig.\,\ref{figure2} together with the simulated evolution of the tilted Bose-Hubbard Hamiltonian, which are in good agreement throughout the experimental ramp.

\subsection{Low-energy properties for frustrated boundary conditions}

\begin{figure}
	\centering
	\includegraphics{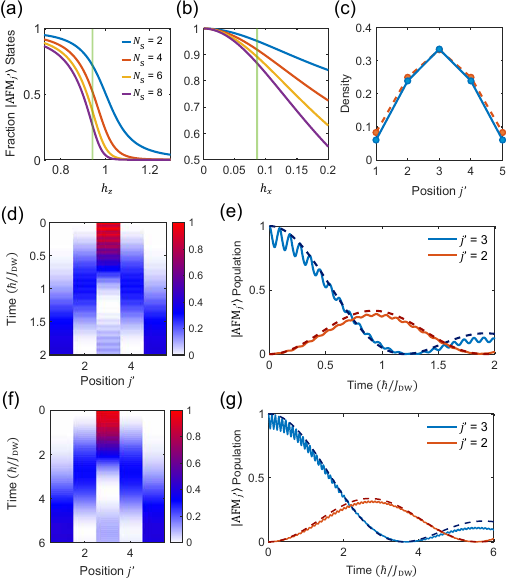}
	\caption{Low-energy properties for frustrated boundary conditions. (a, b) Fraction of domain-wall states $|\textrm{AFM}_{j^\prime} \rangle$ in the ground state of a spin chain for even sizes $N_{\mathrm s}$, as a function of $h_z$ (a) and $h_x$ (b). In (a), the transverse field $h_x = 0.086$ is chosen to match our experimental parameters, and the green line indicates the QCP. In (b), $h_z = 0.73$ is chosen to match the smallest longitudinal field realized experimentally, and the green line indicates $h_x = 0.086$. (c) Local density of domain walls (blue solid line) for a chain of size $N_{\mathrm s} = 10$. The red dashed line shows the density profile of a single particle on five lattice sites with open boundary conditions. (d) Simulated quantum walk of a domain wall for a chain of size $N_{\mathrm s} = 10$, starting from the initial state $|\textrm{AFM}_{j^\prime = 3} \rangle$. (e) Domain-wall density as a function of time, on the central link $(j^\prime = 3)$ and the adjacent link $(j^\prime = 2)$. We also show the analytical prediction for a single-particle quantum walk with tunneling amplitude $J_{\textrm{DW}}$. (f) Simulated quantum walk of a domain wall in the Bose-Hubbard model with $L = 9$ sites at unity filling, starting from the Fock state associated to $ |\textrm{AFM}_{j^\prime = 3} \rangle$. (g) Fitting to Bessel functions yields a reduced tunneling amplitude $J_{\textrm{Fit}} / J_{\textrm{DW}}= 0.33(1)$ compared to Eq.\eqref{eq:J_DW}. In (c-g), simulation parameters have been chosen such that $h_x = 0.086$ and $h_z$ = 0.73.}
	\label{FigureS:DomainWalls}
\end{figure}

In this paragraph, we describe the ground state of the spin model \eqref{eq:Ising} with fixed boundary conditions (i.e.~in the presence of static $|{\uparrow}\rangle$--spins at both ends of the chain) deep in the AFM regime $0 < h_z < 1$. We first discuss the case $h_x = 0$ for which the Hamiltonian \eqref{eq:Ising} is diagonal in the $\{ |{\uparrow}\rangle, |{\downarrow}\rangle\}$--product basis. For an odd number of spins $N_{\mathrm{s}}$, the Hamiltonian has a single ground state given by the perfect AFM ordering $|{\uparrow} {\downarrow} ... {\downarrow} {\uparrow} \rangle$, while the gap to the next excited level is given by the energy cost $J ( 1 - h_z )$ of flipping one of the $|{\downarrow}\rangle$--spins, which is of order $J$ as shown in Fig.\,\ref{FigureS:Gap}. The system is hence gapped in the thermodynamic limit, and the addition of a non-zero transverse field $h_x$ does not qualitatively modify this picture.

For an even number of spins $N_{\mathrm s}$, the ground state becomes degenerate, and each state of the degenerate manifold contains exactly one domain wall made of adjacent $|{\uparrow}\rangle$--spins. There are exactly $N_{\mathrm{DW}} = N_{\mathrm s} / 2$ such configurations which we denote $|\textrm{AFM}_{j^\prime} \rangle_{1 \le j^\prime \le N_{\mathrm{DW}}}$, such that $j^\prime$ indicates that the corresponding domain wall is located on the link between the spins $2 j^\prime - 2$ and $2 j^\prime - 1$ (labels $0$ and $N_{\mathrm s} - 1$ refer to the fixed external spins). Ground state degeneracy occurs because the local antiferromagnetic order cannot propagate from one end of the chain to the other, hence the terminology of frustrated boundary conditions \cite{2015Owerre, 2020Maric, 2021Maric}. These domain walls can also be interpreted as stable excitations located on every other link of the spin chain, with zero tunneling amplitude. A small $h_x > 0$ lifts this degeneracy and restores tunneling for the domain walls. At second order in perturbation theory in $h_x$, the domain-wall tunneling amplitude is given by
\begin{equation}
\label{eq:J_DW}
J_{\textrm{DW}} = \frac{J h_x^2}{4} \left( \frac{1}{h_z} + \frac{1}{1 - h_z} \right).
\end{equation}
The low-energy spectrum of this chain is hence made of $N_{\mathrm{DW}}$ states in a narrow band of width $J_{\textrm{DW}} \ll J$, such that the system is gapless in the thermodynamic limit, as suggested by Fig.\,\ref{FigureS:Gap}. This spectrum is approximately the same as the band structure of a single particle with tunneling amplitude $J_{\textrm{DW}}$.

To illustrate the previous discussion, we plot in Fig.\,\ref{FigureS:DomainWalls}(a, b) the fraction of domain-wall states $|\textrm{AFM}_{j^\prime}\rangle$ present in the ground state for even sizes $N_{\mathrm s}$, as a function of $h_z$ and $h_x$. In addition, we compare in Fig.\,\ref{FigureS:DomainWalls}(c) the local density of domain walls in the ground state with the local density of a single particle with tunneling amplitude $J_{\textrm{DW}}$. As a dynamical signature of this low-energy spectrum, we numerically simulate the quantum walk of a single domain wall prepared at the center of the chain, which we compare with the expected behavior for a single-particle quantum walk \cite{2015Preiss}, see Fig.\,\ref{FigureS:DomainWalls}(d, e). Finally, we discuss the applicability of the previous results to the tilted Bose-Hubbard system \eqref{eq:H} that we realize experimentally. Using numerical simulation of the Bose-Hubbard Hamiltonian with parameters matching our experiments, we time-evolve the Fock state equivalent to a centered domain wall. In Fig.\,\ref{FigureS:DomainWalls}(f, g), we observe a behavior similar to that of a single-particle quantum walk, although with a reduced effective tunneling amplitude compared to Eq.\,\eqref{eq:J_DW}. We attribute this mismatch to the imperfect mapping between the Bose-Hubbard parameters and the spin parameters away from the resonance $E = U$.

\subsection{Influence of disorder}

\begin{figure}
	\centering
	\includegraphics{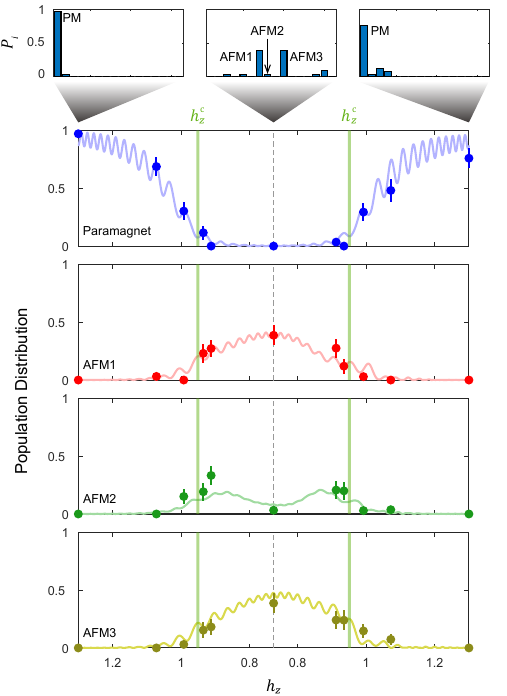}
	\caption{Influence of disorder for frustrated boundary conditions. Evolution of the population distributions for a Bose-Hubbard chain with length $L = 5$ (which corresponds to the system size of $N_\mathrm{s} = 6$ spins) is illustrated. The position of the QCP expected in the thermodynamic limit is shown as a green vertical line. Solid lines are computed by numerical integration of the Schr\"odinger equation, assuming $h \times (5, 0, 3, 5, 0)$~Hz on-site offset potentials at each lattice site $j$ in addition to the tilted Bose-Hubbard model~\eqref{eq:H}. Experimental data points have been reported from Fig.\,\ref{figure4}(b) of the main text. Histograms show population distributions $P_i$ of all possible spin configurations, including those of paramagnetic (PM) and 3 different antiferromagnetic (AFM) ones introduced in Fig.\,\ref{figure4}(a).}
	\label{figure7}
\end{figure}

We numerically explore the influence of on-site potential disorder on the population distributions of the domain-wall states of a chain with the frustrated boundary conditions in the AFM regime. As an illustration, we show in Fig.\,\ref{figure7} that potential disorder at a-few-Hz level is enough to explain the behavior observed in chain of length $N_{\mathrm{s} } = 6$. This level of disorder is consistent with the intrinsic disorder present in our optical lattice, but can also be caused by our projected DMD box potential \cite{2011Simon_n_Greiner, 2016Zupancic_n_Greiner}. In Fig.\,\ref{figure4} of the main text, the shaded areas were obtained by sampling random spatial distributions of the longitudinal field $h_z$ corresponding to uncorrelated potential disorder with a Gaussian distribution of 1 Hz standard deviation per lattice site, i.e. $0.2\%$ of $U$.

\subsection{Spin observables}

\begin{figure}
	\centering
	\includegraphics{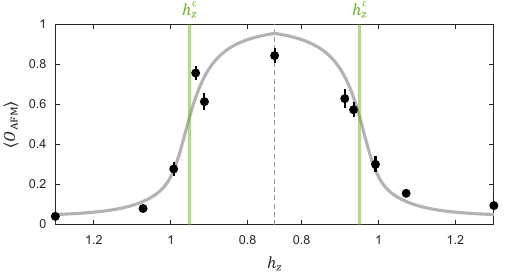}
	\caption{Evolution of the N\'eel order parameter for a Bose-Hubbard chain with length $N_{\mathrm{s} } = 7$ across the QCP. The position of the QCP expected in the thermodynamic limit is shown as a green vertical line. The solid line shows the prediction of exact diagonalization calculations without any free parameters. Error bars denote the 1$\sigma$ statistical errors.}
	\label{FigureS:Order_Parameter}
\end{figure}

\begin{figure}
	\centering
	\includegraphics{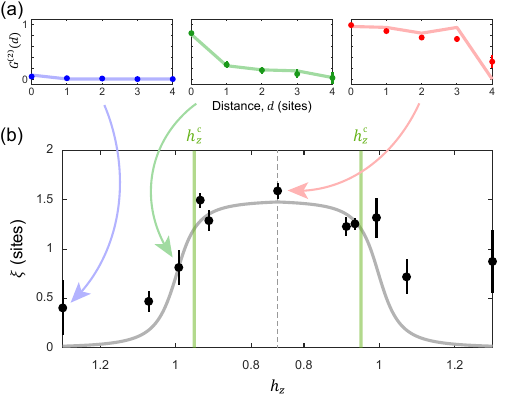}
	\caption{(a) Two-point correlation function $G^{(2) }(d)$ as a function of distance $d$ for the $N_{\mathrm{s} } = 7$ system for various $h_z$ during the first half of the sweep. The spins form little correlation in the paramagnetic regime, while they show long-range correlation in the AFM regime. (b) Evolution of the correlation length $\xi$ as a function of $h_z$. The position of the QCP expected in the thermodynamic limit is shown as a green vertical line. The solid lines show the prediction of exact numerical calculations without any free parameters. Error bars denote the 1$\sigma$ statistical errors, obtained using bootstrap in (b).}
	\label{FigureS:Correlation}
\end{figure}

In the main text, we consider the mean magnetization and its variance
\begin{eqnarray}
	\langle M_z \rangle &=& \left \langle \frac{1}{N_{\mathrm{s} }^*} \sum_{j = 1}^{N_{\mathrm{s} }^*} \sigma_j^z \right \rangle \\
	\langle \left( \delta M_z \right)^2 \rangle &=& \left \langle \left ( \frac{1}{N_{\mathrm{s} }^*} \sum_{j = 1}^{N_{\mathrm{s} }^*} \sigma_j^z \right )^2 \right \rangle - \langle M_z \rangle^2,
	\label{eq:mag}
\end{eqnarray}
where we introduced the Pauli $z$-operator $\sigma^{z}_j$. Here, the observables are estimated for $N_{\mathrm{s} }^* = N_{\mathrm{s} } - 2$ number of dynamic spins, excluding the boundary ones from this calculation. In Fig.\,\ref{FigureS:Order_Parameter}, we also estimate the N\'eel order parameter
\begin{equation}
	\langle O_{\mathrm{AFM} } \rangle = \left \langle \left [\frac{1}{N_{\mathrm{s} }^*} \sum_{j = 1}^{N_{\mathrm{s} }^*} (-1)^j \sigma_j^z \right ]^2 \right \rangle,
	\label{eq:neelO}
\end{equation}
which remains close to 0 when $h_z$ lies in the PM regime and rises to 1 in the AFM regime. Finally, we determine the correlations between the $z$-component of the spins as a function of their distance $d$
\begin{eqnarray}
	\label{eq:G2}
	G^{(2) }(d) &=& \Bigg | \frac{1}{N_{\mathrm{s} }^* - d} \sum_{j = 1}^{N_{\mathrm{s} }^* - d} \langle \sigma_j^z \sigma_{j + d}^z \rangle - \left (\frac{1}{N_{\mathrm{s} }^* - d} \sum_{j = 1}^{N_{\mathrm{s} }^* - d} \langle \sigma_j^z \rangle \right ) \nonumber \\
	&& \left ( \frac{1}{N_{\mathrm{s} }^* - d} \sum_{j = 1}^{N_{\mathrm{s} }^* - d} \langle \sigma_{j + d}^z \rangle \right ) \Bigg |,
\end{eqnarray}
where the above definition accounts for the non-homogeneity of the chain due to the fixed boundary conditions. The behavior of $G^{(2) }(d)$ is shown in Fig.~\ref{FigureS:Correlation}(a).

In the thermodynamic limit $N_{\mathrm s} \rightarrow +\infty$, the correlation function for this Ising system decays exponentially with a correlation length $\xi$. For our relatively small system sizes, fitting the correlation function to an exponential decay is not robust and we rather estimate the correlation length via the following relation
\begin{equation}
	\xi = \frac{\sum_{d=0}^{N_{\mathrm{s} }^*} d\,G^{(2) }(d)}{\sum_{d=0}^{N_{\mathrm{s} }^*} G^{(2) }(d)}.
	\label{eq:xi}
\end{equation}
As shown in Fig.~\ref{FigureS:Correlation}(b), the correlation length in the PM regime is close to 0, while it peaks at a finite value in the AFM regime.

\subsection{Single-site von Neumann entropy}

\begin{figure}
	\centering
	\includegraphics{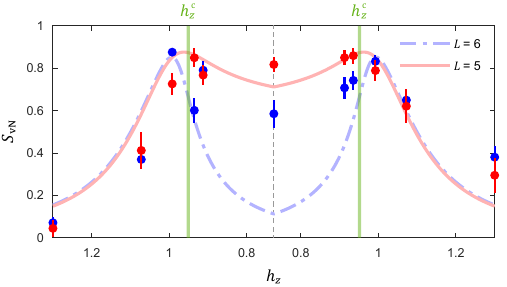}
	\caption{Single-site von Neumann entropy in the $L = 6$ (i.e.~$N_{\mathrm{s} } = 7$, blue) and $L = 5$ (i.e.~$N_{\mathrm{s} } = 6$, red) systems. For $L = 6$, the entropy shows a maximum at the QCP before decreasing towards its initial level when rewinding the ramp. For $L = 5$, the entropy remains at its highest level even in the AFM regime. The position of the QCP expected in the thermodynamic limit is shown as a green vertical line. The blue and red lines denote the prediction of exact numerical calculations without any free parameters. Error bars denote the 1$\sigma$ statistical errors obtained using bootstrap.}
	\label{FigureS:Entropy}
\end{figure}

In this paragraph, we use the single-site von Neumann entropy to investigate the build-up of entanglement across the spin chain \cite{2009Horodecki}. The entanglement entropy of a subsystem can be accessed after partially tracing out the density matrix of the whole system.  Experimentally, the single-spin density matrix cannot be directly determined from full-counting statistics, due to presence of coherences between the $|{\uparrow}\rangle$ and $|{\downarrow}\rangle$ spin states. Conversely, the single-site density matrix is diagonal in the atom number basis $n_j = \{0, 1, 2\}$ due to the constraint on the conserved total particle number (super-selection rules) \cite{2016Kaufman_n_Greiner, 2019Lukin_n_Greiner}. For a given lattice site $j$, we define the single-site von Neumann entropy
\begin{equation}
	S_{\textrm{vN}, j} = - \sum_{n_j = 0}^2 p_{n_j} \log p_{n_j},
\end{equation}
such that $p_{n_j}$ is the probability for the lattice site $j$ to contain $n_j$ atoms. Fig.\,\ref{FigureS:Entropy} shows the evolution of the spatial average of the single-site entropy, $S_{\textrm{vN}} = \frac{1}{L} \sum_{j = 1}^L S_{\textrm{vN}, j}$, as a function of ramp time. For $L = 6$ ($N_{\mathrm{s} } = 7$), the entanglement entropy shows a clear maximum around the QCP, and decreases towards its initial level at the end of the inverse ramp. The increase in entropy therefore comes mainly from the entanglement build-up within the chain \cite{2019Blatt_n_Zoller}. On the other hand, for $L = 5$ ($N_{\mathrm{s} } = 6$), the entropy remains high even in the AFM regime due to the superposition of domain wall states already mentioned in Fig.\,\ref{figure4}, and then decreases back to a lower level after the inverse ramp.

\end{document}